\documentclass[twocolumn]{revtex4}
\usepackage{graphicx}
\usepackage[usenames]{color}
\usepackage{times}
\usepackage[english]{babel}

\usepackage[ansinew]{inputenc}
\usepackage{amsmath,amssymb,bm}
\usepackage{mathrsfs}
\usepackage[all,line]{xy}
\usepackage[T1]{fontenc}
\usepackage{verbatim}
\usepackage{ifthen}
\usepackage{verbatim}

\newcommand{\af}[1]{\!\left(#1\right)}
\newcommand{\tuborg}[1]{\left\{ #1 \right\}}
\newcommand{\kantpar}[1]{\left[ #1 \right]}

\renewcommand{\vec}[1]{\mathbf{#1}}                       
 
\newcommand{\mathlow}[1]{_{\mathrm{#1}}}
\newcommand{\mathhigh}[1]{^{\mathrm{#1}}}

\newcommand{\Exp}[1]{e^{#1}}
\newcommand{\Op}[1]{\hat #1}

\newcommand{\Tot}{\mathlow{tot}}
\newcommand{\Res}{\mathlow{res}}

\newcommand{\aop}{\Op{a}^\dagger}
\newcommand{\aned}{\Op{a}}

\newcommand{\defi}{\equiv}

\newcommand{\kL}{k_L}

\newcommand{\Atom}{a}
\newcommand{\Molecule}{m}
\newcommand{\Type}{\sigma}
\newcommand{\SubAtom}{_\Atom}
\newcommand{\SubMolecule}{_\Molecule}
\newcommand{\SubType}{_\Type}

\newcommand{\SpinUp}{\uparrow}
\newcommand{\SpinDown}{\downarrow}
\newcommand{\SupSpinUp}{^{\SpinUp}}
\newcommand{\SupSpinDown}{^{\SpinDown}}

\newcommand{\NumSites}{M}

\newcommand{\NumTot}{N\Tot}
\newcommand{\NumU}{\NumTot\SupSpinUp}
\newcommand{\NumD}{\NumTot\SupSpinDown}
\newcommand{\NumA}{N\SubAtom}
\newcommand{\NumAU}{N\SubAtom\SupSpinUp}
\newcommand{\NumAD}{N\SubAtom\SupSpinDown}
\newcommand{\NumM}{N\SubMolecule}

\newcommand{\Energi}{E}

\newcommand{\E}{\Energi}

\newcommand{\EAnum}[1]{\Energi\SubAtom^{#1}}
\newcommand{\EMnum}[1]{\Energi\SubMolecule^{#1}}
\newcommand{\Einum}[1]{\Energi\SubType^{#1}}

\newcommand{\KemPot}{\mu}

\newcommand{\KemPotM}{\mu\SubMolecule}
\newcommand{\KemPotAU}{\mu\SubAtom\SupSpinUp}
\newcommand{\KemPotAD}{\mu\SubAtom\SupSpinDown}

\newcommand{\FF}{\eta}

\newcommand{\Eres}{\E\Res}

\newcommand{\EresDis}{\E\mathlow{dis}}

\newcommand{\EF}{E\mathlow{F}}

\newcommand{\ERi}{E_{R,\Type}}
\newcommand{\ERA}{E_{R,\Atom}}
\newcommand{\ERM}{E_{R,\Molecule}}

\newcommand{\TF}{T\mathlow{F}}

\newcommand{\f}{f}
\newcommand{\fA}{\f\SubAtom}
\newcommand{\fM}{\f\SubMolecule}
\newcommand{\fType}{\f\SubType}

\newcommand{\dens}{\rho}
\newcommand{\dA}{\dens\SubAtom}
\newcommand{\dM}{\dens\SubMolecule}

\newcommand{\kB}{k\mathlow{B}}

\newcommand{\refEq}[1]{(\ref{#1})}

\newcommand{\term}[1]{#1}
\newcommand{\Pot}{V}
\newcommand{\RelNumM}{\chi}
\newcommand{\EresDisLeft}{\EresDis^{-}}
\newcommand{\EresDisRight}{\EresDis^{+}}
\newcommand{\EresDisLeftRight}{\EresDis^{\pm}}
\newcommand{\LatticePot}{L}

\renewcommand{\figurename}{Figure}
\newcommand{\sectionname}{section}

\def\ket#1{\mathinner{|{#1}\rangle}}
\def\braket#1{\mathinner{\langle{#1}\rangle}}

{\catcode`\|=\active 
  \gdef\Braket#1{\begingroup
     \ifx\SavedDoubleVert\relax
       \let\SavedDoubleVert\|\let\|\BraDoubleVert
     \fi
     \mathcode`\|32768\let|\BraVert
     \left<{#1}\right>\endgroup}
}
\def\BraVert{\@ifnextchar|{\|\@gobble}
     {\egroup\,\mid@vertical\,\bgroup}}
\def\BraDoubleVert{\egroup\,\mid@dblvertical\,\bgroup}
\let\SavedDoubleVert\relax

{\catcode`\|=\active
  \gdef\set#1{\mathinner{\lbrace\,{\mathcode`\|32768\let|\midvert #1}\,\rbrace}}
  \gdef\Set#1{\left\{
     \ifx\SavedDoubleVert\relax \let\SavedDoubleVert\|\fi
     \:{\let\|\SetDoubleVert
     \mathcode`\|32768\let|\SetVert
     #1}\:\right\}}
}
\def\midvert{\egroup\mid\bgroup}
\def\SetVert{\@ifnextchar|{\|\@gobble}
    {\egroup\;\mid@vertical\;\bgroup}}
\def\SetDoubleVert{\egroup\;\mid@dblvertical\;\bgroup}

\begingroup
 \edef\@tempa{\meaning\middle}
 \edef\@tempb{\string\middle}
\expandafter \endgroup \ifx\@tempa\@tempb
 \def\mid@vertical{\middle|}
 \def\mid@dblvertical{\middle\SavedDoubleVert}
\else
 \def\mid@vertical{\mskip1mu\vrule\mskip1mu}
 \def\mid@dblvertical{\mskip1mu\vrule\mskip2.5mu\vrule\mskip1mu}
\fi

\begin{document}
\title{Statistical mechanics of a Feshbach coupled Bose-Fermi gas in an optical lattice}
\author{O. S{\o}e S{\o}rensen}
\affiliation{Lundbeck Foundation Theoretical Center for Quantum System Research, Department of Physics and Astronomy, University of Aarhus, DK-8000 {\AA}rhus C, Denmark}
\author{P. B. Blakie}
\affiliation{Jack Dodd Centre for Quantum Technology, Department of Physics, University of Otago, New Zealand}
\author{N. Nygaard}
\affiliation{Lundbeck Foundation Theoretical Center for Quantum System Research, Department of Physics and Astronomy, University of Aarhus, DK-8000 {\AA}rhus C, Denmark}

\begin{abstract}
  We consider an atomic Fermi gas confined in a uniform optical lattice potential, where the atoms can pair into molecules via a magnetic field controlled narrow Feshbach resonance. The phase diagram of the resulting atom-molecule mixture in chemical and thermal equilibrium is determined numerically in the absence of interactions under the constraint of particle conservation. In the limiting cases of vanishing or large lattice depth we derive simple analytical results for important thermodynamic quantities. One such quantity is the dissociation energy,  defined as the detuning of the molecular energy spectrum with respect to the atomic one for which half of the atoms have been converted into dimers. Importantly we find that the dissociation energy has a non-monotonic dependence on lattice depth.  
\end{abstract}

\maketitle

\section{Introduction}

The use of Feshbach resonances with ultracold atoms in optical lattices provides a new avenue for creating molecules. 
This system has seen experimental realizations in a variety of systems with bosonic and fermionic species \cite{Stoferle2006,Kohl2006,Chin2006,Thalhammer2006,Ospelkaus2006}.
For the case of fermionic atoms the Feshbach assisted conversion is to molecular bosons with strikingly different behavior in the degenerate  regime.
In this article we study the statistical mechanics of such a degenerate gas of fermionic atoms confined in an optical lattice and subject to a Feshbach resonance. Our interest is the chemical equilibrium between the fermionic atoms and their dimerized, bosonic counterparts.

For an ultracold atomic gas a Feshbach resonance is observed when a closed channel bound state is coupled to the scattering continuum of an energetically open channel. The position of this resonance is tunable, since the different channels correspond to different combinations of internal atomic states and hence experience different Zeeman shifts in an applied magnetic field. 
As the closed channel bound state is tuned from above to below the open channel threshold, the resonance becomes a true molecular bound state of the two-body system~\cite{Kohler2006}. 
The parameter controlling the position of the resonance as the magnetic field is varied is the resonance energy $\Eres$, which is the detuning of the closed channel bound state from the open channel threshold. 

Previously the thermodynamics of an atomic Fermi gas with Feshbach resonant atom-molecule conversion has been studied both in free space~\cite{Carr2004} and in a harmonic confining potential~\cite{Williams2004,Watabe2007}. For $\Eres<0$ the dimers are stable against dissociative decay and hence are real molecules. For $\Eres>0$ the dimers have a finite lifetime, but we can still discuss chemical equilibration of the gas in the steady state limit, where a detailed balance is established between molecule formation and decay. In this respect we also use the term molecules for the unstable dimer states, which in an ensemble will have a finite occupation at any instant. Note that this implies a narrow Feshbach resonance, for which the scattering resonance is a consequence of a long-lived quasi-stationary state embedded in the continuum~\cite{Gurarie2007}. In the structured continuum of the optical lattice the molecular states may also be stable against dissociative decay at energies above the continuum threshold, if their energies lie in the band gaps~\cite{Syassen2007,Winkler2006,Nygaard2008a,Nygaard2008b}.

For a narrow Feshbach resonance it has been established that the existence of a small parameter facilitates a \emph{quantatively} correct description of the many-body physics based on a pertubative expansion~\cite{Gurarie2007, Gurarie2004}. Building on this insight, we consider for simplicity an ideal gas mixture, such that the only effect of the Feshbach resonance is to maintain the chemical equilibrium between the two species. Even this simple model captures the essential physics of atom-molecule conversions in experiments~\cite{Williams2006}, and it is exact in the limit where the resonance is infinitey narrow~\cite{Gurarie2007}. With a finite atom-molecule coupling an effective atom-atom interaction arises upon the elimination of the molecular degrees of freedom. We emphasize that even though this effective interaction may diverge as the magnetic field is varied across the resonance, all thermodynamic quantities such as the chemical potential and the condensate fraction remain well behaved (see e.g.~\cite{Ohashi2002, Kokkelmans2002}). In particular, for a narrow Feshbach resonance the ideal gas thermodynamics remains qualitatively correct when the atom-molecule coupling is small but finite~\cite{Gurarie2007}. 

We believe that the optical lattice introduces considerable new physics, and that the best way to understand the resulting changes in the thermodynamics is to consider the ideal case first. Thus, for a narrow Feshbach resonance we anticipate that the inclusion of interactions will only impact our results quantitatively. Notwithstanding, for a broad Feshbach resonance interactions play a crucial role, and hence in that case the thermodynamics of the BCS-BEC crossover requires more advanced modelling than presented here, i.e. a full many-body theory for resonantly interacting Fermi atoms~\cite{Haussmann2007, Haussmann2008, Hu2006a, Hu2006b, He2007, Perali2004}. For a deep lattice this has been studied at zero temperature by Koetsier et al.~\cite{Koetsier2006}. However, even for a broad resonance a thermodynamic description based only on the molecular binding energy, the temperature and the total number of atoms can give a molecule fraction which agrees well with experimental data across the Feshbach resonance~\cite{Chin2004}.

The thermal properties of ultracold atoms in optical lattices have received considerable recent attention. In the degenerate regime ideal Bose and Fermi gases have been investigated in the uniform lattice \cite{Blakie2004a,Blakie2005a} and in the presence of additional external harmonic confinement \cite{Kohl2006a,Blakie2007a,Blakie2007b}. Interactions play an important role in deep lattices, and more recent work has examined the effect these interactions have on the thermal excitation generated during the preparation of bosonic Mott-insulating states \cite{Ho2007a,Gerbier2007a}, and on the feasibility of achieving the fermionic Neel state \cite{Koetsier2008a}. Finite temperature mixtures of atomic Bose and Fermi gases in lattices have been studied \cite{Cramer2008a} in an attempt to explain recent experiments \cite{Ospelkaus2006a}.

Our system has several parameters which can be varied independently in a numerical calculation. We only show results for half filling which is of the greatest interest in relation to current experiments. However, we have found that systems with filling fractions less than or equal to unity have much the same behavior. For filling fraction larger than unity the excited bands play a larger role, and some of the conclusions presented here have to be amended. For simplicity we restrict our analysis to the case of a spin-balanced Fermi gas. There are many interesting effects associated with spin polarization, but these do not play a major role in the transition between atoms and molecules which we intend to study here. Instead the essential quantities governing the phase diagram are the resonance energy, i.e. the energy offset between atoms and molecules, the lattice depth, and the temperature. Tuning the resonance energy at a fixed lattice depth shifts the energy spectra of the atoms and the molecules with respect to each other, thereby moving the point of chemical equilibrium. Since the atoms and molecules experience different lattice potentials, the depth of the periodic potential determines where the transition between the two species occurs.

\section{Formalism} \label{sec:Thermodynamics}
We consider a simple cubic optical lattice with $\NumSites^3$ sites containing $\NumTot$ identical fermionic atoms of mass $m\SubAtom$ with two internal states labeled $\ket{\SpinUp}$ and $\ket{\SpinDown}$, which we will refer to as spin-up and spin-down respectively. In this paper we restrict our attention to the case of an equal number of atoms in the two internal states
\begin{align}
  \NumU = \NumD = \frac{\NumTot}{2} ,
\end{align}
and we define the \term{filling fraction} $\FF$ to be the average number of each kind of atom per site
\begin{align}
  \FF \defi \frac{\NumU}{\NumSites^3} = \frac{\NumD}{\NumSites^3} = \frac{\NumTot}{2\NumSites^3} .
\end{align}

With the Feshbach resonance a spin-up and a spin-down atom can couple to a bosonic dimer with mass $m\SubMolecule = 2 m\SubAtom$. In the following we determine the phase diagram of the system under the assumption that the only interaction between the atoms is the Feshbach resonance, which maintains chemical equilibrium between unbound atoms and diatomic molecules. Since we are assuming thermodynamic equilibrium, it is then meaningful to define the number of molecules $\NumM$ and the number of atoms $\NumA$ as the average number of molecules and atoms, respectively, in a long time period where all the external parameters are kept constant. At all times particle conservation leads to
\begin{align}
  \NumTot = \NumA + 2\NumM = \NumAU + \NumAD + 2\NumM ,
  \label{eq:NumCons}
\end{align}
where $\NumAU$ and $\NumAD$ are the numbers of unbound spin-up and spin-down atoms respectively.

In thermal equilibrium the temperatures of the atoms and the molecules are the same, $T$, while the condition of chemical equilibrium can be expressed from the atomic and molecular chemical potentials as
\begin{align}
  \KemPotM = \KemPotAU + \KemPotAD.
\end{align}
Since, in the spin-balanced case, the spin-up and spin-down atoms have identical thermal behavior, we have $\KemPotAU = \KemPotAD \defi \KemPot$ and therefore $\KemPotM = 2\KemPot$. 

\begin{figure*}[t]  
  \includegraphics[width=0.8\textwidth]{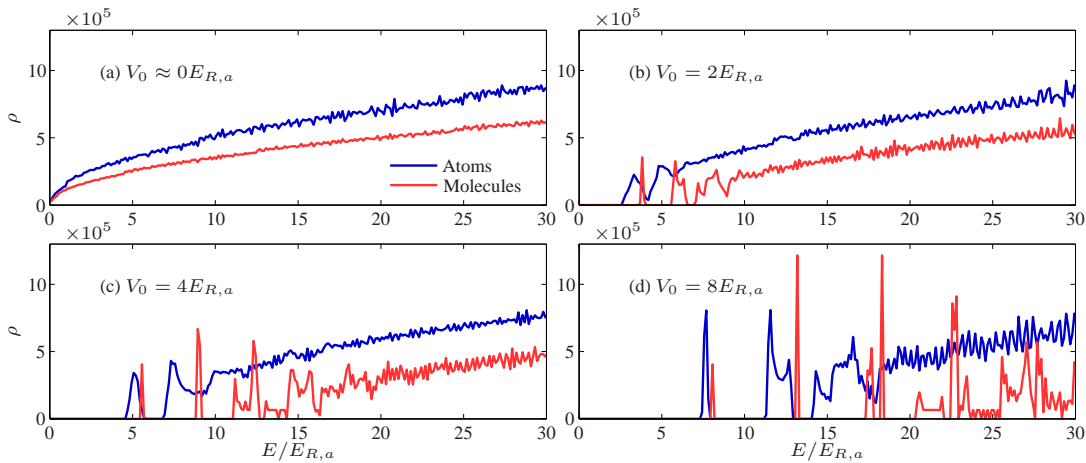}
  \caption{(Color online) The density of states of a 3D optical lattice calculated by binning the energy levels into intervals of length $\Delta E = 0.13 \ERA$. The blue (dark gray) and red (light gray) curves show the density of the atomic and molecular states, respectively.} 
  \label{fig:Density}
\end{figure*}

Our choice of energy convention is to measure all single particle states from the lattice potential zero of energy, and furthermore, we subtract the  magnetic field shift $\Eres$ from the molecular states. Thus the distribution functions take the form 
\begin{subequations}
  \label{eq:fsigma}
  \begin{align}
    \fA \af{E} &\defi
    \frac{1}{\Exp{(E- \KemPot)/\kB T} + 1} ,
    \label{eq:fA}
    \\
    \fM \af{E} &\defi
    \frac{1}{\Exp{(E+ \Eres - 2\KemPot)/\kB T} - 1}.
    \label{eq:fM}
  \end{align}
\end{subequations}
As we discuss in further detail below $\EAnum{r}$ and $\EMnum{r}$ are the available single particle energy states for atoms and molecules, respectively, where $r=0,1,2,\ldots$, with $r=0$ corresponding to the ground state. Importantly, the molecular chemical potential is bounded from above by the lowest possible single-molecule energy
\begin{align}
  \KemPotM \leq \EMnum{0} + \Eres,
  \label{eq:muupbound}
\end{align}
since the molecules are governed by Bose-Einstein statistics.
The density of states for atoms and molecules are given by
\begin{subequations}
  \begin{align}
    \label{DOS_A}
    \dA\af{E} &= 2 \sum_{r=0}^{\infty} \delta\af{E - \EAnum{r}}, \\
    \dM\af{E} &= \sum_{r=0}^{\infty} \delta\af{E - \EMnum{r}},
  \end{align}
\end{subequations}
where the factor of two in (\ref{DOS_A}) arises because there are two types of atoms with identical energy spectra. The total number of atoms and molecules for given values of $\Eres$, $\KemPot$ and $T$ can be found by integrating the density of states weighted by the occupation  over all energies. Because the energy levels are discrete, this becomes a sum
\begin{subequations}
  \label{eq:NumPart}
  \begin{align}
    \NumA &= \int dE\,  \dA\af{E} \fA \af{E}  =2\sum_{r}\fA \af{\EAnum{r}} ,  \\
    \NumM &= \int dE\,  \dM\af{E} \fM \af{E}  =\sum_{r} \fM \af{\EMnum{r}}.
  \end{align}
\end{subequations}
Given the energy levels and the total particle number the chemical potential is determined at any temperature and resonance energy by the constraint (\ref{eq:NumCons}).

\subsection{Energy levels in the lattice} \label{sec:E_in_lattice}
We consider a simple cubic optical lattice created by the overlap of three orthogonal pairs of counterpropagating lasers with wavelength $\lambda_L = 2\pi/\kL$. This gives rise to the potential
\begin{align}
  \Pot\SubType\af{\vec{x}} &= \Pot_{0,\Type} \LatticePot(\vec{x}) ,
\end{align}
for the particle type $\Type$ (corresponding to atoms ``$\Atom$'' or molecules ``$\Molecule$''), where $\LatticePot$ is the dimensionless shape of the potential
\begin{align}
  \label{eq:latpot}
  \LatticePot(\vec{x})&= \kantpar{\sin^2\af{\kL x} + \sin^2\af{\kL y} + \sin^2\af{\kL z}} .
\end{align}
The molecules experience a lattice potential twice as deep as that in which the atoms move, $\Pot_{0,m} = 2\Pot_{0,a}\defi 2\Pot_0$, since the Stark shift of a molecule is the sum of those for each atom.

Now the problem is to understand how the difference in particle mass and apparent lattice depth  for atoms and molecules affects the behavior of their respective single particle states, obtained by solving the time-independent Schr\"odinger equation
\begin{align}
  \tuborg{-\frac{\hslash^2\nabla^2}{2m\SubType} + \Pot_{0,\Type}\LatticePot(\vec{x})} \psi\SubType^r(\vec{x}) = \Einum{r} \psi\SubType^r(\vec{x}) .
  \label{Shrodinger3D}
\end{align}
There is a simple relationship between the atomic and molecular spectra, which is revealed by
transforming  (\ref{Shrodinger3D}) for each species into their respective recoil energies, $\ERi \defi \hslash^2 \kL^2 / 2m\SubType$:
\begin{align}
  \tuborg{-\frac{\nabla^2}{k_L^{2}} + \bar{\Pot}_{0,\Type} \LatticePot(\vec{x})} \psi\SubType^r(\vec{x}) = \bar{E}\SubType^r\psi\SubType^r(\vec{x}),
  \label{Shrodinger3Db}
\end{align}
where barred quantities are in recoil units. The advantage of Eq.~(\ref{Shrodinger3Db}) is that, since the left hand side operator only depends on $\Type$ via the quantity $\bar{\Pot}_{0,\sigma}$,  the spectrum is of the form
\begin{align}
  \bar{E}\SubType^r &= \bar{E}^r(\bar{\Pot}_{0,\Type}).
  \label{eq:barE}
\end{align}

The relationship between $\bar{\Pot}_{0,a}$ and $\bar{\Pot}_{0,m}$ is quite simple: the molecules are twice as heavy as the atoms and see a lattice that is twice as deep as the one experienced by the atoms. As a result the atomic and molecular potentials are related as
\begin{align}
  \bar{\Pot}_{0,\Molecule}=4\bar{\Pot}_{0,\Atom},
  \label{Vbar}
\end{align}
i.e. in the respective recoil units the molecules see a lattice that is four times deeper. In general, an optical lattice with depth exceeding one recoil unit has a considerable effect on the spectral properties of the confined particles. Therefore, the difference in lattice depth for the atoms and the molecules has a rather profound effect on the properties of our system.

We note that our choice to define the atomic energy origin as the lattice potential zero, rather than the atomic groundstate, is to emphasize the differential lattice confinement effects on the atoms and the molecules.

\section{Numerics} 
Because we are considering a separable potential~(\ref{eq:latpot}), the 3D spectrum of the time-independent Schr\"odinger equation is most efficiently calculated via the 1D eigenvalues, which are easily determined by numerical diagonalization.  For the data presented here we have used a lattice with $31\times 31\times 31$ sites; this is sufficiently large that this system can be regarded as being approximately in the thermodynamic limit. We only exemplify results for $\FF = \frac{1}{2}$ since these represent the generic behaviour of the systems with a filling fraction less than unity.

We numerically determine  the chemical potential as a function of $T$ and $\Eres$ under the condition of conservation of the total number of particles, \refEq{eq:NumCons}. For a given point $(\KemPot , \Eres , T)$ in the phase diagram any thermodynamic quantity, such as the energy or the entropy, may then be calculated as the sum of an atomic and a molecular contribution.

\subsection{Density of states}\label{DOSnumerics}
The exact 3D energy levels can be found by making all possible combinations of three 1D energy levels. However, for the purposes of calculation the number of individual states needed is unwieldy, and it is desirable to  construct a \emph{binned}  density of states by gathering the energy levels in small energy intervals and representing them by the centre of the interval. Doing this we obtain the graphs shown in \figurename~\ref{fig:Density}. The reliability of the binning procedure is confirmed in \figurename~\ref{fig:Density}(a), where we recover the usual $\sqrt{E}$ dependence of the densities of states in the limit of vanishing lattice depth. By comparing {\figurename}s \ref{fig:Density}(b) and (d) one sees that the molecular density of states for $\Pot_0 = 2 \ERA$ and the atomic density of states for $\Pot_0 = 8 \ERA$ have the exact same shape, though the latter is scaled on both axes by a factor of two (see Eqs. (\ref{eq:barE}) and (\ref{Vbar})). To simplify the numerical task without losing effects due to details in the energy spectrum we have used the exact energy levels for the lowest energy bands only and the binned density of states for the higher energy levels. The justification for this approximation is that the high energy domain is only relevant if the temperature is high, in which case the distributions $\fType\af E$ are sufficiently slowly varying that we can consider them constant over small energy intervals. 

As the lattice depth is increased, band gaps emerge in the 3D density of states due to the gaps in the 1D energy spectrum. With reference to results shown in \figurename~\ref{fig:Density} we make the following observations of the spectral properties:
\begin{enumerate}
\item The molecular density of states has clear characteristics of a being in a much deeper lattice than the atomic system, i.e. for a given $\Pot_0$ the molecular spectrum has smaller band widths and larger band gaps than the atomic system. Most importantly, even for shallow lattices (e.g. $V_0=2\ERA$), the molecule ground band is very narrow compared to the atomic ground band. These observations are consistent with the discussion below Eq.~(\ref{Vbar}).  
\item  There is a positive offset in the ground state energy of the molecules relative to the atom states. As shown in Appendix \ref{sec:app}, this shift arises from anharmonic effects in the lattice, and for the deep lattice limit it is given by
\begin{align}
  \EMnum{0}-\EAnum{0} \approx \frac{3}{8}\af{1 + \frac{3}{8} \frac{1}{\sqrt{\Pot_{0,\Atom}/\ERA}}} \ERA.
  \label{EMEAdiff}
\end{align}
Such shifts should be measurable in experiments as a displacement of the magnetic field position where half of the atoms have been converted to molecules in an adiabatic sweep across the Feshbach resonance (see \sectionname~\ref{sec:DissociationEnergy}).
\end{enumerate}

\section{Phase diagram} \label{sec:PhaseDiagram}
The phase diagram is characterized by the molecule fraction, defined as 
\begin{align}
  \RelNumM\af{\Eres,T,\Pot_0 , \FF} \defi \frac{2\NumM}{\NumTot}
\end{align}
ranging from zero (all atoms) to unity (all molecules). In this section we first examine the molecule fraction as a function of the position of the resonance and the temperature of the system. We then consider how the low temperature chemical equilibrium is affected by changing the lattice depth.

A convenient energy scale  is the Fermi energy $\EF\af{\FF, \Pot_0}$, which is taken to be the highest occupied energy level, when all the atoms are unbound. We note that for our choice of energy origin the relevant Fermi temperature for characterizing degeneracy is given by $\TF = (\EF - \EAnum{0} )/ \kB$, with $\EAnum{0}$ the atomic ground state energy.

\subsection{Properties at fixed lattice depth} \label{sec:FixedDepth}
\begin{figure}[tbh]
  \includegraphics[width = \columnwidth]{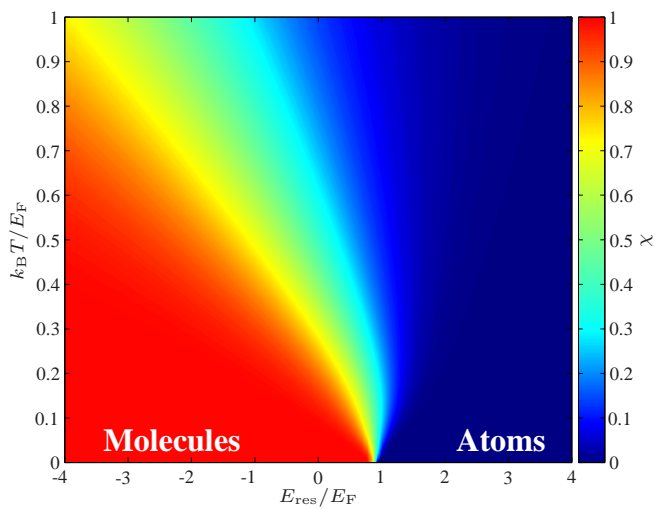}
  \caption{(Color online) Molecule fraction in a $(\Eres , T)$ phase diagram for $\Pot_0 = 5 \ERA$ and $\FF = \frac{1}{2}$.}
  \label{fig:AntalKont1}
\end{figure}

The molecular fraction is shown in \figurename~\ref{fig:AntalKont1} as a function of $\Eres$ and $T$ for a fixed lattice depth. Those results show that tuning the Feshbach resonance, i.e. $\Eres$, provides a direct way of varying the composition of the gas. For large and positive values of $\Eres$, the molecular states are at much higher energy than the atomic states and are not thermally accessible, thus realizing a pure atomic gas. Conversely, for large and negative values of $\Eres$, the system exists as a pure molecular gas. At low temperature the transition between these two limiting regimes occurs  when $\Eres$ is close to $\EF$ (see \figurename~\ref{fig:AntalKont1}). We remark that for a free system or a gas trapped in a harmonic potential the transition from atoms to molecules starts at $\Eres \approx 2 \EF$~\cite{Williams2004, Gurarie2007}.

\begin{figure*}[tb]
  \centering
  \includegraphics[width=0.8\textwidth]{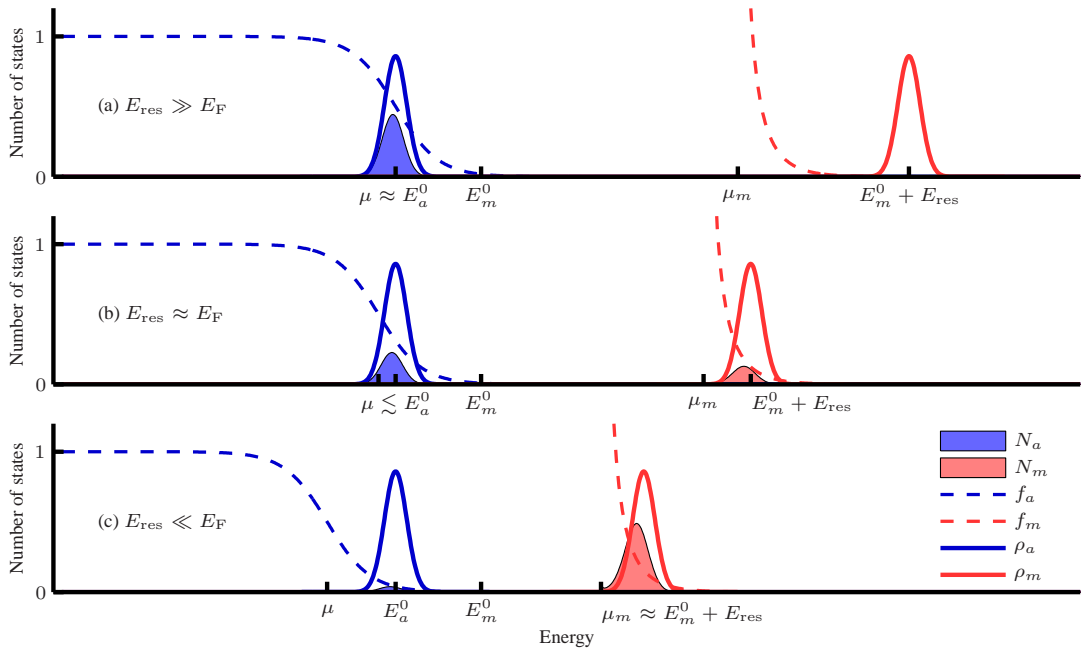}
  \caption{(Color online) Schematic illustration of how $\KemPot$ changes when the resonance energy is varied across $\EF$. For clarity we consider a deep lattice where the density of states of the lowest atomic- and molecular band is a narrow peak as illustrated by the solid lines. The distributions $\fA\af E$ and $\fM\af E$ are indicated by the dashed lines, and the shading under the peaks indicates the population of the atomic and molecular levels. Note that the $\Eres$-dependence is included in the position of the peak in the molecular density of states and \emph{not} in the Bose-Einstein distribution.}
  \label{fig:Overgang}
\end{figure*}

The characteristic asymmetric fan-shape of the $\RelNumM$-contours occurs because when the temperature increases, molecules will dissociate as two seperate atoms are entroptically favorable over a single molecule. Thus if we follow vertical lines in the phase diagram in \figurename~\ref{fig:AntalKont1} towards higher temperatures, the fraction of molecules must decrease eventually. 

The chemical potentials for degenerate Bose and Fermi gases exhibit markedly different behavior, the former being constrained by the lowest available single particle level, while the latter depends explicitly on the number of particles in the system. Furthermore, in our system chemical equilibrium constrains the molecular chemical potential to be twice that of the atoms. Thus it is clear that the change in atomic and molecular populations induced by varying the applied magnetic field must be accompanied by a change in the behavior of the chemical potential. This can be described qualitatively by considering how the atoms and molecules are distributed over their respective energy levels when $\Eres$ is swept across the Feshbach resonance as sketched in \figurename~\ref{fig:Overgang}. We identify three separate regimes:
\begin{description}
\item[(a) Pure Fermi gas limit, $\bm{\Eres \to\infty}$:]
  If $\Eres$ is sufficiently high, the lowest molecular energy level $\EMnum{0} + \Eres$ is nearly unoccupied and all atoms are unbound. We then have
  \begin{align}
    \KemPot \approx \EF  \quad \text{for} \quad T \approx 0.
  \end{align}
  This corresponds to \figurename~\ref{fig:Overgang}(a) where the narrow energy bands are sketched as peaks.
  
  For filling fraction $\FF \leq 1$ the Fermi energy lies in the ground band. In the limit of a deep lattice the lowest band becomes sufficiently narrow that we can make the approximation $\EF \approx \EAnum{0}$ (see Appendix \ref{sec:app}).

\item[(b) Intermediate region, $\bm{\Eres \approx \EF}$:]
 When $\EMnum{0} + \Eres$ approaches $\KemPotM$ from above, molecules will start to form while the number of atoms decreases as shown in \figurename~\ref{fig:Overgang}(b). When $\Eres$ decreases further, $\KemPotM$ also has to decrease due to the condition~(\ref{eq:muupbound}). Since $\KemPotM = 2\KemPot$, the atomic chemical potential therefore also starts to  decrease.

\item[(c) Pure Bose gas limit, $\bm{\Eres \to -\infty}$:] 
  Finally, when the resonance energy becomes low enough, $\KemPot$ has decreased so much that almost no atomic states are occupied, and the result is a nearly pure molecular gas corresponding to \figurename~\ref{fig:Overgang}(c). In this limit we have $\KemPotM \approx \EMnum{0} + \Eres$ at low temperatures, i.e.,
  \begin{align}
    \KemPot =  \frac{\EMnum{0} + \Eres}{2} \quad \text{for} \quad T \approx 0,
    \label{BoseGasLimit}
  \end{align}
\end{description}
From this description it is clear that the transition from atoms to molecules is more abrupt the narrower the lowest energy bands are.

\begin{figure}[tb]
  \includegraphics[width=\columnwidth]{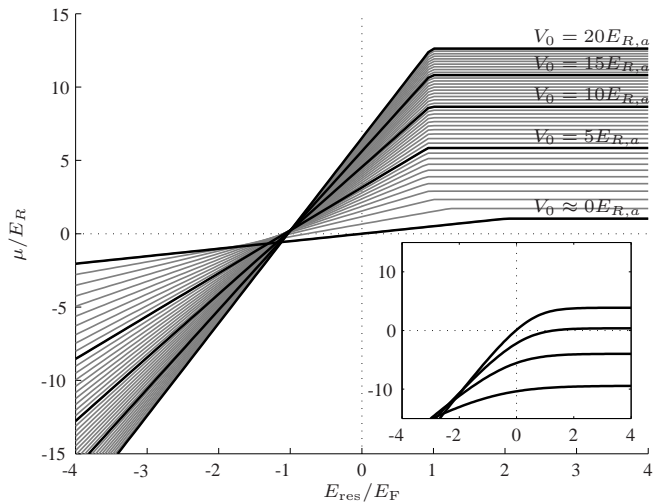}
  \caption{Chemical potential as a function of $\Eres$ for $\FF = \frac{1}{2}$ and different potential depths in the limit $T\approx 0$. This corresponds to following a horizontal line in the bottom of phase diagrams like in \figurename~\ref{fig:AntalKont1}. The linear behaviour in the left side is characteristic for the pure Bose gas regime, and the difference in the slopes arises because the Fermi energy varies with the potential depth. Inset: chemical potential curves for $\Pot_0 = 5$, $10$, $15$, and $20 \ERA$ (from bottom to top) for a finite temperature of $ 5.6 \ERA/\kB$.}
  \label{fig:KemPotLavT}
\end{figure}

The low temperature behaviour of $\KemPot$ at half filling is shown in \figurename~\ref{fig:KemPotLavT} for a range of different potential depths and the effects discussed above are evident. Linear fits to $\KemPot$ in the molecular regime turns out to be in excellent agreement with the pure Bose gas limit estimate (\ref{BoseGasLimit}) and  the $\KemPot$-plateaus in the pure Fermi gas limit reveal the dependence of $\EF$ on the lattice depth. Note that the arguments above do not tell us anything about the behaviour in the transition zone between the constant and the linear regime, but merely that at sufficiently high $\Eres$ the chemical potential must be constant, while for $\Eres$ tuned sufficiently below the resonance the chemical potential is linear. The fact that the transitions between these two regimes is so sharp as in \figurename~\ref{fig:KemPotLavT} is a consequence of the very low temperature. However, the general behaviour of plateaus on the atomic side of the resonance and a linear variation of $\KemPot$ on the molecular side remain also at higher temperatures in the limits $\Eres \to \pm \infty$ as illustrated in the inset of \figurename~\ref{fig:KemPotLavT}.

\section{Dissociation energy} \label{sec:DissociationEnergy}
A noticeable feature in \figurename~\ref{fig:KemPotLavT} is that the transition point where the atoms start to form molecules varies with the potential depth. In the absence of the optical lattice potential it lies at $\Eres\approx 2 \EF$, as is well-known from BCS-BEC cross-over theories for a narrow Feshbach resonance~\cite{Gurarie2007}. As the lattice depth is increased the resonance energy corresponding to the onset of the transition first decreases and then increases slightly towards a limiting value of $\Eres\approx \EF$ in the deep lattice limit. We will elaborate on these effects in the following and explain the behavior for deep lattice potentials.

To quantify the location of the chemical transition we define the \term{dissociation energy} $\EresDis$ as the value of the resonance energy where half of the atoms have been converted to molecules, $\RelNumM(\EresDis) = \frac{1}{2}$~\cite{Williams2004}. In addition, we introduce the quantities $\EresDisLeftRight$, delineating the zero temperature conversion zone, such that no unbound atoms are found for $\Eres<\EresDisLeft$ and no molecules exist for $\Eres>\EresDisRight$. The lower limit is defined by the condition
\begin{align}
  \EMnum{0} + \EresDisLeft = 2\EAnum{0},
  \label{eq:EdisLeftEstimate}
\end{align}
which specifies where the first molecules start to break up as the resonance energy is increased from the molecular side of the resonance. The other end of the transition zone, where the first molecules are formed as the resonance energy is decreased starting with a pure atomic gas, is defined by the condition that the lowest molecular level passes twice the energy of the highest occupied atomic state, i.e.
\begin{align}
  \EMnum{0} +  \EresDisRight = 2\EF .
  \label{eq:EdisRightEstimate}
\end{align}

\subsection{Free space limit} \label{sec:nolattice}
These estimates simplify in the limit $\Pot_0 = 0$ where the lowest atomic and molecular energy levels are only separated by $\Eres$, i.e. $\EAnum{0} = \EMnum{0}=0$. We then have
\begin{align}
  \begin{array}{rl}
    \EresDisLeft &= 0 \\
    \EresDisRight &= 2\EF
  \end{array}
  \qquad \text{for}  \qquad \Pot_0= 0 .
\end{align}
To find $\EresDis$ for a vanishing potential we note that in this case the density of states has a squareroot dependence on the energy
\begin{align}
  \dA(E) = C \sqrt{E}
\end{align}
with $C=\pi M^3/2\ERA^{3/2}$.The following argument applies at zero temperature: the number of unbound fermionic atoms, found  by integrating the atomic density of states from zero to the chemical potential, is $\NumA=\frac{2}{3}C\KemPot^{3/2}$. If $\KemPot = \EF$, this equals the total number of atoms in the system, and in general we obtain the expression
\begin{align}
  \RelNumM
  &= 1 - \frac{\NumA}{\NumTot}
  = 1 - \af{\frac{\KemPot}{\EF}}^{3/2},
\end{align}
for the molecule fraction (at $T=0$). In the transition zone $\Eres = 2 \KemPot$ (since $\EMnum0=0$), such that
\begin{align}
  \RelNumM = 1 - \af{\frac{\Eres}{2\EF}}^{3/2},
\end{align}
and the dissociation energy in this limit then follows from the condition $\RelNumM = \frac{1}{2}$:
\begin{align}
  \EresDis = 2^{1/3} \EF \approx 1.26 \EF \quad \text{for} \quad \Pot_0 = 0.
  \label{eq:EdislowV0}
\end{align}
This is confirmed by the numerical calculations, see \figurename~\ref{fig:EDis}. Similar arguments can be used to show that $\EresDis = 2^{2/3} \EF$ for a gas trapped in a harmonic potential~\cite{Williams2004}. For a harmonic trap one also finds $\EresDisLeft = 0$ and $\EresDisRight = 2 \EF$ since $\EMnum{0} = \EAnum{0} \ll \EF$.

\subsection{Deep lattice limit} \label{sec:deeplattice}
\begin{figure}[t]
  \centering      
  \includegraphics[width=\columnwidth]{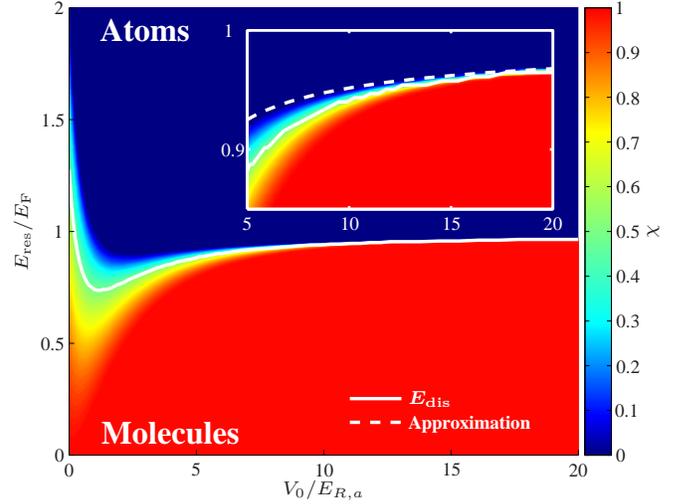}      
  \caption{(Color online) Dissociation energy vs. optical lattice depth. For deep lattice potentials $\EresDis$ can be approximated by Eq. \refEq{eq:EdisTeo} (white line), while the limiting behavior for a vanishing lattice depth is given by Eq. \refEq{eq:EdislowV0} (dashed white line). Note that since the Fermi energy increases with increasing lattice depth, the scaling on the vertical axis depends on $\Pot_0$.}
  \label{fig:EDis}
\end{figure}

In the opposite limit of a deep optical lattice the 3D energy bands become increasingly narrow and $\EresDisLeft$ and $\EresDisRight$ approach each other. It is therefore reasonable to estimate $\EresDis$ to lie halfway between them, and inserting (\ref{eq:EdisLeftEstimate}) and (\ref{eq:EdisRightEstimate}) yields
\begin{align}
  \EresDis &\to \EF - (\EMnum{0} - \EAnum{0}) \qquad \text{for} \qquad \Pot_0 \to \infty.
\end{align}
In the deep lattice limit the narrow ground band is well-described by a tight-binding approach (see Appendix \ref{sec:app}), which gives an analytic expression for $\EMnum{0} - \EAnum{0}$ (\ref{eq:Enuldiff}). We then have that
\begin{align}
  \EresDis &\approx \EF - \frac{3}{8}\af{1  + \frac{3}{8} \frac{1}{\sqrt{\Pot_{0,\Atom} / \ERA}}}\ERA.
  \label{eq:EdisTeo}
\end{align}
This estimate, valid for a deep lattice, is plotted in \figurename~\ref{fig:EDis}, where we also show the numerical halfway mark for the atom-molecule conversion, $\EresDis$, and there is good agreement between the numerical results and \refEq{eq:EdisTeo} in the deep lattice limit. Note that since $\EF \approx \EAnum{0}$ for a deep lattice, (\ref{eq:Enulapprox}) indicates that in this limit $\EF / \ERA \approx 3\sqrt{\Pot_0 / \ERA}$ and hence that $\EresDis$ eventually approaches $\EF$ as the lattice depth increases
\begin{align}
  \EresDis \to \EF \quad \text{for} \text \quad \Pot_0 \to \infty,
\end{align}
since the first term in \refEq{eq:EdisTeo} is then the dominant one. The deviation from this limit in a deep lattice is given by the anharmonic corrections in (\ref{eq:EdisTeo}).

\subsection{Lattice induced resonance shift} \label{sec:estimate}
These results demonstrate that the lattice induced energy shift between atomic and molecular degrees of freedom alters the dissociation condition for the Feshbach resonance. In an adiabatic sweep of the magnetic field across the resonance the effect of the lattice is to modify the molecule formation curve ($\RelNumM$ vs. $B$). In particular, the magnetic field corresponding to conversion of half of the atoms to molecules, $B^{\Pot_0}\mathlow{dis}$, is shifted with respect to its value in the absence of the optical lattice, $B^{0}\mathlow{dis}$, by an amount $\delta B^{\Pot_0}\mathlow{dis}= B^{\Pot_0}\mathlow{dis}-B^{0}\mathlow{dis}$, which in the deep lattice limit may be estimated from
\begin{align}
  \delta B^{\Pot_0}\mathlow{dis}
  &\approx \frac{\EresDis^{\Pot_0}-\EresDis^0}{\Delta\mu}\approx\frac{\EF^{\Pot_0}-1.26\EF^0}{\Delta\mu}.
\end{align}
Here we have used that $\Eres$ varies linearly with $B$ with a slope $\Delta\mu$ given by the magnetic moment difference between the open and the closed channels. The Fermi energy with no lattice is $\EF^0 = \af{\frac{6 \FF}{\pi}}^{2/3} \ERA$ and for deep lattices we can approximate $\EF^{\Pot_0} \approx 3 \sqrt{\Pot_0 \ERA}$. Inserting this we get
\begin{align}
  \delta B^{V_0}\mathlow{dis}
  &\approx\frac{3 \sqrt{\frac{\Pot_0}{\ERA}} - 1.26\af{\frac{6 \FF}{\pi}}^{2/3}} {\af{\frac{\Delta\mu}{\mu\mathlow{B}}} \af{\frac{m_a}{\mathrm{amu}}} \af{\frac{\lambda\mathlow{L}}{\mathrm{nm}}}^2} \cdot {1.43 \times 10^{-5}} \mathrm{G} .
\end{align}
For $^6$Li in a lattice with wavelength $\lambda\mathlow{L}=1032 \ \mathrm{nm}$ and taking $\Delta \mu = 2\mu\mathlow{B}$ we find that $B^{\Pot_0}\mathlow{dis}-B^{0}\mathlow{dis}\approx {0.11} \ \mathrm{G}$ for a lattice depth of $\Pot_0 = 15 \ERA$ assuming a filling fraction of $\FF = \frac{1}{2}$. This is comparable to the magnetic field width of the resonance $\Delta B= {0.23} \ \mathrm{G}$ at resonance at ${543.26} \ \mathrm{G}$~\cite{Strecker2003}. A similar expression has been found experimentally for the lattice induced shift of the magnetic position where the Feshbach molecule enters the continuum~\cite{Syassen2007}.

\section{Conclusion}
We have presented results for the thermodynamics of an atom-molecule mixture in an optical lattice potential under the condition of chemical equilibrium adjusted by a Feshbach resonance which controls the energy offset between the atomic and molecular levels.

The phase diagram has been determined and we have analyzed the behavior of the chemical potential and the fraction of molecules with emphasis on the low-temperature regime. In particular we have identified the dissociation energy defined as the energy offset where $50\%$ of the atoms have been dimerized into molecules. The chemical conversion takes place in a transistion zone of the magnetic field controlled resonance energy, which narrows and shifts as the lattice depth is increased and the Bloch bands approach discrete energy levels. Furthermore, the dissociation energy shifts with the lattice depth, as the molecules and the free atoms experience different lattice potentials, and hence their energy levels are displaced with respect to one another as the lattice depth is changed. In the deep lattice limit an analytic, precise expression for the center of the transistion zone was obtained by treating anharmonic corrections to the energy levels in the lattice wells perturbatively.

Our results thus show that in an optical lattice potential the position and width of the Feshbach resonance, as indicated by the conversion of atoms into dimers and vice versa, depends on the lattice depth. We have shown that the lattice induced shift should be measurable for an atom with a narrow resonance such as $^6$Li. 

By focussing on the ideal gas case we have been able to clarify the effect of the uniform optical lattice potential on the atom-molecule equilibrium. The inclusion of interactions and an inhomogenous trap potential will be areas of future development to completely describe this system and make quantatative comparison with experiments possible.

\acknowledgments{N. N. acknowledges financial support by the Danish Natural Science Research Council.} 

\appendix 
\section{The deep lattice limit}
\label{sec:app}
When the optical lattice is very deep, we can make a tight-binding approximation and regard the potential as a collection of $\NumSites^3$ independent wells.
The lowest energy levels are then those for one well with a degeneracy equal to the number of sites. A Taylor expansion of the 1D potential ($V_{0,\Type}\sin^2(k_Lx)$) at $x = 0$ to 6th order gives
\begin{align}
  \Pot\SubType\mathhigh{1D}\af{x}
  \approx \Pot_{0,\Type} \af{\af{\kL x}^2 - \frac{1}{3}\af{\kL x}^4 + \frac{2}{45}\af{\kL x}^6} ,
\end{align}
so each lattice well can be described as a harmonic oscillator with the frequency
\begin{align}
  \omega\SubType = \sqrt{\frac{2{\kL}^2 \Pot_{0,\Type}}{m\SubType}},
\end{align}
with two anharmonic terms
\begin{align}
  \Op H_4 &= - \frac{\Pot_{0,\Type}}{3} \af{\kL \Op x}^4 ,  \\
  \Op H_6 &= \frac{2\Pot_{0,\Type}}{45} \af{\kL \Op x}^6 ,
\end{align}
that we will treat with perturbation theory. The unpertubed energy levels for the harmonic oscillator are
\begin{align}
  \mathcal{E}_{\Type}^{n(0)}
  = \af{n + \frac{1}{2}} \hslash \omega\SubType
  = \af{2n + 1} \sqrt{\Pot_{0,\Type}\ERi},
\end{align}
and the first order corrections
\begin{align}
  \Delta  \mathcal{E}_{\Type,i}^{n(1)} = \braket{n|\Op H_i|n} , \qquad i = 4 , 6
\end{align}
can be calculated by expressing the factor $(\kL \Op x)^i$ in terms of the ladder operators for the harmonic oscillator
\begin{align}
  (\kL \Op x)^i &= \tuborg{\af{\frac{\ERi}{4 \Pot_{0,\Type}}}^{1/4} \af{\aop + \aned}}^{i} .
\end{align}
We thus obtain the anharmonic first order energy shifts
\begin{align}
  \Delta  \mathcal{E}_{\Type,4}^{n(1)} &= - \frac{\ERi}{4} \af{2n^2 + 2n + 1} ,\\
  \Delta  \mathcal{E}_{\Type,6}^{n(1)} &= \frac{1}{36} \frac{\ERi}{\sqrt{\Pot_{0,\Type} / \ERi}} \af{4 n^3 +6n^2 + 8n + 3} ,
\end{align}
but we also need to consider the second order pertubation contributions from $\Op H_4$ to the $n$'th energy level
\begin{align}
  \Delta  \mathcal{E}_{\Type,4}^{n(2)}
  &= \sum_{m \neq n} \frac{|\braket{m |\Op H_4|n}|^2}{ \mathcal{E}_{\Type,2}^{n} -  \mathcal{E}_{\Type,2}^{m}} \nonumber\\
  &= -\frac{1}{288} \frac{\ERi}{\sqrt{\Pot_{0,\Type}/\ERi}} \sum_{m \neq n} \frac{|\braket{m | \af{\aop + \aned}^{4}|n}|^2}{m - n},
\end{align}
since this contribution is of the same order as the first order pertubation contribution from $\Op H_6$. In the general case the sum only gets contributions from $m = n\pm 2$ and $m = n\pm 4$, and in the simplest case $n = 0$, where the contributions from $m = n-4$ and $m = n-2$ vanish, we get
\begin{align}
  \Delta  \mathcal{E}_{\Type,4}^{0(2)}
  &= - \frac{7}{48} \frac{\ERi}{\sqrt{\Pot_{0,\Type}/\ERi}},
\end{align}
resulting in the approximation
\begin{align}
  \mathcal{E}_{\Type}^{0}
  &\approx \af{\sqrt{\frac{\Pot_{0,\Type}}{\ERi}} - \frac{1}{4} - \frac{3}{48} \frac{1}{\sqrt{\Pot_{0,\Type} / \ERi}}}\ERi 
\label{eq:Enulapprox}
\end{align}
for the lowest 1D energy level.

This derivation is valid for both atoms and molecules, and therefore the splitting between $\EAnum{0}$ and $\EMnum{0}$ in three dimensions is given by $\EMnum{0}-\EAnum{0}=3( \mathcal{E}\SubMolecule^{0}- \mathcal{E}\SubAtom^{0})$, i.e.
\begin{align}
  (\EMnum{0} - \EAnum{0})
  &\approx\frac{3}{8} \af{1  + \frac{3}{8} \frac{1}{\sqrt{\Pot_{0,\Atom} / \ERA}}}\ERA
  \label{eq:Enuldiff}
\end{align}
where we have used the relation (\ref{Vbar}) and that $\ERM = \ERA/2$. The last term in (\ref{eq:Enuldiff}) can be made as small as desired by choosing a deep enough lattice potential, but the first term always remains resulting in an unavoidable energy difference between the lowest energy band of the two components. As discussed in \sectionname~\ref{sec:PhaseDiagram} this offset has observable consequences for the onset of molecule production in sweeps of the magnetic field across the resonance.


\begin{thebibliography}{90}

\bibitem{Stoferle2006} T. St{\"o}ferle, H. Moritz, K. G{\"u}nter, M. K{\"o}hl, and T. Esslinger, Phys. Rev. Lett. {\bf 96}, 030401 (2006).
\bibitem{Kohl2006} M. K{\"o}hl, K. G{\"u}nter, T. St{\"o}ferle, H. Moritz, and T.
Esslinger, J. Phys. B: At. Mol. Opt. Phys. {\bf 39}, S47 (2006).
\bibitem{Chin2006} J. K. Chin, D. E. Miller, Y. Liu, C. Stan W. Setiawan, C. Sanner, K. Xu, and W. Ketterle, Nature {\bf 443}, 961 (2006).
\bibitem{Thalhammer2006}  G. Thalhammer, K. Winkler, F. Lang, S. Schmid, R.
Grimm, J. Hecker Denschlag, Phys. Rev. Lett. {\ 96}, 050402 (2006).
\bibitem{Ospelkaus2006} C. Ospelkaus, S. Ospelkaus, L. Humbert, P. Ernst, K. Sengstock, and K. Bongs, Phys. Rev. Lett. {\bf 97}, 120402 (2006).

\bibitem{Kohler2006} T. K{\"o}hler, K. G{\'o}ral, and P. S. Julienne, Rev. Mod. Phys. {\bf{78}}, 1311 (2006).
\bibitem{Carr2004} L. D. Carr, G. V. Shlyapnikov, and Y. Castin, Phys. Rev. Lett. {\bf 92} 150404 (2004).
\bibitem{Williams2004} J. E. Williams, N. Nygaard, and C. W. Clark, New. J. Phys. {\bf 6}, 123 (2004).
\bibitem{Watabe2007} S. Watabe, T. Nikuni, N. Nygaard, J. E. Williams, and C. W. Clark, J. Phys. Soc. Jpn. {\bf 76}, 064003 (2007).

\bibitem{Gurarie2004} A. V. Andreev, V. Gurarie, and L. Radzihovsky, Phys. Rev. Lett. {\bf 93}, 130402 (2004).
\bibitem{Gurarie2007} V. Gurarie and L. Radzihovsky, Ann. Phys. {\bf 322}, 2 (2007).


\bibitem{Winkler2006}  K. Winkler, G. Thalhammer, F. Lang, R. Grimm, J. Hecker Denschlag, A. J. Daley, A. Kantian, H. P. B{\"u}chler, and P. Zoller, Nature (London) {\bf{441}}, 853 (2006).
\bibitem{Nygaard2008a} N. Nygaard, R. Piil, and K. M{\o}lmer, Phys. Rev. A {\bf 77}, 021601(R) (2008).
\bibitem{Nygaard2008b} N. Nygaard, R. Piil, and K. M{\o}lmer, Phys. Rev. A {\bf 78}, 023617 (2008).
\bibitem{Syassen2007} N. Syassen, D. M. Bauer, M. Lettner, D. Dietze, T. Volz, S. D{\"u}rr, and G. Rempe, Phys. Rev. Lett. {\bf{99}}, 033201 (2007). 
\bibitem{Williams2006}   J. E. Williams, N. Nygaard, and C. W. Clark, New J. Phys. {\bf 8}, 150 (2006).

\bibitem{Ohashi2002} Y. Ohashi and A. Griffin, Phys. Rev. Lett. {\bf 89}, 130402 (2002).

\bibitem{Kokkelmans2002} S. J. J. M. F. Kokkelmans, J. N. Milstein, M. L. Chiofalo, R. Walser, and M. J. Holland, Phys. Rev. A {\bf 65}, 053617 (2002).


\bibitem{Haussmann2008} R. Haussmann and W. Zwerger, arXiv:0805.3226v3 (2008).

\bibitem{Hu2006a} H. Hu, X. J. Liu and P. D. Drummond, Phys. Rev. A {\bf 73}, 023617 (2006).

\bibitem{Hu2006b} H. Hu, X. J. Liu and P. D. Drummond, Europhys. Lett. {\bf 74} 574 (2006).


\bibitem{He2007} Y. He, C. C. Chien, Q. Chen, and K. Levin, Phys. Rev. B {\bf 76}, 224516 (2007).

\bibitem{Perali2004} A. Perali, P. Pieri, L. Pisani, and G. C. Strinati, Phys. Rev. Lett. {\bf 92}, 220404 (2004).

\bibitem{Haussmann2007} R. Haussmann, W. Rantner, S. Cerrito and W. Zwerger, Phys. Rev. A {\bf 75}, 023610 (2007). 

\bibitem{Koetsier2006} Arnaud Koetsier, D. B. M. Dickerscheid and H. T. C. Stoof, Phys. Rev. A {\bf 74}, 033621 (2006). 

\bibitem{Chin2004} C. Chin and R. Grimm, Phys. Rev. A {\bf 69}, 033612 (2004). 

\bibitem{Blakie2005a} P. B. Blakie and A. Bezett, Phys. Rev. A {\bf 71}, 033616 (2005).

\bibitem{Blakie2004a} P. B. Blakie and J. V. Porto, Phys. Rev. A {\bf 69}, 013603 (2004). 

\bibitem{Blakie2007b} P. B. Blakie and Wen-Xin Wang,
Phys. Rev. A {\bf 76}, 053620 (2007).

\bibitem{Blakie2007a} P. B. Blakie, A. Bezett, and P. Buonsante,
Phys. Rev. A {\bf 75}, 063609 (2007).

\bibitem{Kohl2006a}Michael K{\"o}hl, Phys. Rev. A {\bf 73}, 031601(R) (2006).  


\bibitem{Ho2007a} Tin-Lun Ho and Qi Zhou, Phys. Rev. Lett. {\bf 99}, 120404 (2007).

\bibitem{Gerbier2007a} Fabrice Gerbier, Phys. Rev. Lett. {\bf 99}, 120405 (2007).

\bibitem{Koetsier2008a} Arnaud Koetsier, R.A. Duine, Immanuel Bloch, and H.T.C. Stoof, Phys. Rev. A {\bf 77}, 023623 (2008).

\bibitem{Cramer2008a} M. Cramer, S. Ospelkaus, C. Ospelkaus, K. Bongs, K. Sengstock, and J. Eisert,  Phys. Rev. Lett. {\bf 100}, 140409 (2008).

\bibitem{Ospelkaus2006a} S. Ospelkaus, C. Ospelkaus, O. Wille, M. Succo, P. Ernst, K. Sengstock, and K. Bongs, Phys. Rev. Lett. {\bf 96}, 180403 (2006)

\bibitem{Strecker2003} K. E. Strecker, G. B. Partridge, and R. G. Hulet, Phys. Rev. Lett. {\bf 91}, 080406 (2003).

\end{thebibliography}
\end{document}